\documentclass[conference]{IEEEtran}
\usepackage{amssymb}
\IEEEoverridecommandlockouts
\usepackage{cite}
\usepackage{amsmath,amssymb,amsfonts}
\usepackage{amsthm}
\usepackage{graphicx}
\usepackage{textcomp}
\usepackage{xcolor}
\usepackage{algorithm}
\usepackage{algorithmicx}
\usepackage[noend]{algpseudocode}
\algrenewcommand\textproc{}
\usepackage{amsmath}
\usepackage[caption=false, font=footnotesize]{subfig}
\usepackage{multirow}

\newcommand{\todo}[1]{\textcolor[rgb]{1.0, 0, 0}{#1}}

\theoremstyle{definition}
\newtheorem{definition}{Definition}

\def\BibTeX{{\rm B\kern-.05em{\sc i\kern-.025em b}\kern-.08em
		T\kern-.1667em\lower.7ex\hbox{E}\kern-.125emX}}
\begin{document}
	
	\title{
	Index-Based Scheduling for Parallel State Machine Replication
		\thanks{Gang Wu is supported by the NSFC (Grant No. 61872072) and the State
Key Laboratory of Computer Software New Technology Open Project Fund (Grant
No. KFKT2018B05).}
	}
		\author{\IEEEauthorblockN{Gang Wu\textsuperscript{1,2}, Guodong Zhao\textsuperscript{1}, Yidong Song\textsuperscript{1}}
		\IEEEauthorblockA{\textsuperscript{1}School of Computer Science and Engineering, Northeastern University - Shenyang, China\\
		\textsuperscript{2}State Key Laboratory for Novel Software Technology, Nanjing University - Nanjing China\\
		wugang@mail.neu.edu.cn, gdzhao@stumail.neu.edu.cn, ydsong@stumail.neu.edu.cn}
			}
	
	\maketitle
	
	\begin{abstract}

		State Machine Replication (SMR) is a fundamental approach to designing service with fault tolerance. However, its requirement for the deterministic execution of transactions often results in single-threaded replicas, which cannot fully exploit the multicore capabilities of today's processors. Therefore, parallel SMR has become a hot topic of recent research. The basic idea behind it is that independent transactions can be executed in parallel, while dependent transactions must be executed in their relative order to ensure consistency among replicas. The dependency detection of existing parallel SMR methods is mainly based on pairwise transaction comparison or batch comparison. These methods cannot simultaneously guarantee both effective detection and concurrent execution. Moreover, the scheduling process cannot execute concurrently, which introduces extra scheduling overhead as well. In order to further reduce scheduling overhead and ensure the parallel execution of transactions, we propose an efficient scheduler based on a specific index structure. The index is composed of a Bloom Filter and the associated transaction queues, which provides an efficient dependency detection and preserve necessary dependency information respectively. Based on the index structure, we further devise an elaborated concurrent scheduling process. The experimental results show that the proposed scheduler is more efficient, scalable and robust than the comparison methods.

	\end{abstract}
	
	\begin{IEEEkeywords}
		fault tolerance, state machine replication, high performance, distributed systems
	\end{IEEEkeywords}
	
	\section{Introduction}
Large-scale online service systems need to ensure  high availability and high efficiency of the services.
State Machine Replication (SMR) \cite{lamport1978time,schneider1990implementing} based on various consensus protocols, such as Paxos\cite{lamport1998part} and PBFT\cite{castro2002practical}, is a common approach to designing fault-tolerate services.
According to SMR model, even some of the replicas fail, the services will be kept available with the consistent replicas.
SMR achieves strong consistency\cite{herlihy1990linearizability} by regulating every replica executing the same transactions in the same order:
(i) every available replica receives all the same transactions eventually;
(ii) all replicas must agree on the same order of the transactions received; and
(iii) every replica starts from the same initial state and executes the agreed transactions deterministically (ie., transaction must guarantee ACID and the transaction's changes to the state of the records are a function of only the initial state of the records and the transaction itself).

	
As we know, SMR is mainly designed to improve the system's availability rather than its performance\cite{hunt2010zookeeper,ongaro2014search,verma2015large,shvachko2010hadoop,ousterhout2010case}.
The requirement of the sequential execution of the total order (same order on all replicas) transactions makes it difficult for SMR to take full advantage of multi-core servers.
It cannot directly execute transactions concurrently because the uncertainty of thread scheduling and lock competition would result in the undeterministic execution.
However, the sequential execution is not a necessary requirement for consistency \cite{schneider1990implementing}.
In short, dependent transactions(access the same records) must be handled in the same relative order on each replica to keep consistency, while independent transactions(access the different records) can be executed in parallel, which can fully utilize the processor's multi-core processing ability.
Thus, basing on transaction semantics, how to use the transaction independence to improve the performance of SMR has become a hot research direction \cite{kotla2004high,evekapritsos2012all,marandi2014rethinking,alchieri2018early,7967165,alchieri2017reconfiguring,8671580}.

For example, CBASE \cite{kotla2004high} is a classic parallel replication framework proposed to enhance the performance of PBFT algorithm.
It sets up a scheduler for every replica which constructs a dependency graph by finding the dependencies \emph{\textbf{pairwise}} among transactions in their total order.
Based on the dependency graph, the scheduler dispatches transactions to idle threads in the thread pool for execution.
Once a transaction is executed by one thread, the scheduler removes it from the graph and responds to clients.
The scheduler of CBASE maximizes concurrency among executions while ensuring replica consistency.

However, recent research \cite{7967165} has shown that, under the conditions of high workload, due to the high overhead of pairwise transaction comparisons, determining dependencies among transactions that have not yet been executed is a performance bottleneck.
To overcome this problem, batchCBASE\cite{7967165} determine the dependencies by \emph{\textbf{batch}} comparison rather than a single transaction comparison once a time, which greatly reduces the times of comparison dramatically. However, it increases the possibility of inter-batch dependencies, and as transactions in each batch are executed sequentially, it loses some of the parallelism for those transactions within a batch.
In this way, batchCBASE provides a possible trade-off between the parallel execution and dependency detection.
Moreover, in order to promise replica consistence and operation safety, the scheduling process of CBASE and batchCBASE are in single-threaded mode, which means the scheduler and worker threads cannot access the dependency graph at the same time, it introduce more overhead to the system.

In summary, parallel SMR schedulers now face four challenges:
1) faster detection of transaction dependencies;
2) not sacrificing any parallelism of the execution;
3) concurrent scheduling process; and
4) ensuring correctness.
In this paper, we propose an efficient scheduler based on a specific index structure to address the above challenges.
It consists of a special Bloom filter and corresponding transaction queues for each filter element, with the Bloom filter, the dependencies among transactions can be detected within a constant time.
Transaction queues can maintain the total order relations of the transactions and also simplify the representation of the transaction dependency graph.
Moreover, the proposed scheduler supports record-granularity locks with the help of the above mentioned index structure, thereby supporting the concurrent scheduling process(specifically the \emph{insert}, \emph{remove}, and \emph{get} operations) of transactions.
In summary, the proposed method can efficiently solve the performance loss problem caused by the heavy scheduling overhead from the dependency graph based comparisons, and it can guarantee the execution parallelism under various workloads with different dependency rates.
To show the proposed model’s advantages in throughput, scalability and robustness in comparison with CBASE and batchCBASE, experiments are conducted and analyzed on a database prototype.
Furthermore, the consistency among replicas and other scheduling safety propositions are proved formally.


The remainder of this paper is organized as follows.
The system model is described in section \ref{Sec:SystemModel}.
The parallel SMR model of CBASE and batchCBASE are introduced in Section \ref{Sec:ParallelSMR}.
In Section \ref{Sec:OurApproach}, the proposed index-based scheduling approach is described in detail.
The experimental results are shown in Section \ref{Sec:Experiments}.
Finally, we introduce some related work in Section \ref{Sec:RelatedWork} and conclude in Section \ref{Sec:Conclusion}.

\section{System Model}\label{Sec:SystemModel}
We assume a general distributed service system model, which is composed of an unbounded client sets $C=\{c_1,c_2,...\}$ and a bounded server set $S=\{s_1,s_2,...,s_n\}$.
All servers in $S$ are replicas of each other and work together to provide highly available services to the clients where the Paxos protocol is used to ensure consistency.
The message transmission among distributed replicas is in asynchronous mode, which allows arbitrary message loss and delay.
We assume that replicas follow the fail-stop model and never encounter a Byzantine error, which means the state of each replica is either \emph{correct} or \emph{crash}, and hence the system with $2f+1$ replicas can tolerate $f$ replicas crashing simultaneously.

The system ensures that if a request message $m$ is sent without failing, all the unfaulty replicas will receive it, and eventually $m$ will be decided in the consensus instance $i$, which is called that the replica accepts $(i,m)$.
The Paxos protocol can promise that at least half of the replicas will accept $(i,m)$, and no replica will accept $(i,\hat{m})$ or $(\hat{i},m)$, where $m \neq \hat{m}$ and $i\neq \hat{i}$.
Intuitively, all messages exist in most replicas or in none of them.
If the messages exist, the order of messages on each replica is exactly the same, ie., \emph{total order}.
Every replica must handle messages in the total order.

\begin{figure}[htbp]
		\centering
		\subfloat[Standard SMR]{        \label{fig-standard}
			\begin{minipage}[c]{\linewidth}
				\centering
				\centerline{\includegraphics[scale=0.35]{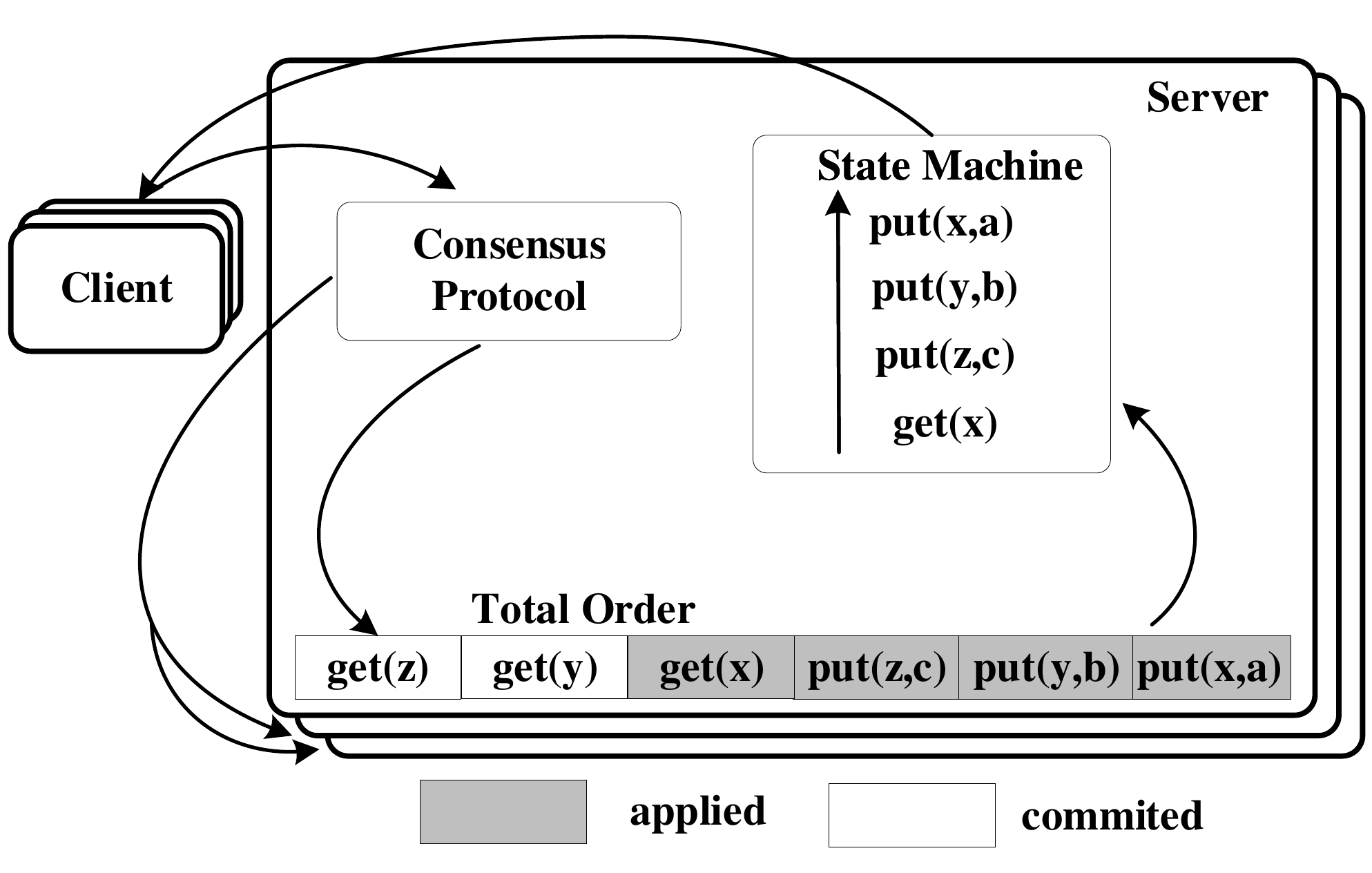}}
				
			\end{minipage}
		}
		\
		\subfloat[Parallel SMR]{      \label{fig-parallel}
			\begin{minipage}[c]{\linewidth}
				\centering
				\centerline{\includegraphics[scale=0.35]{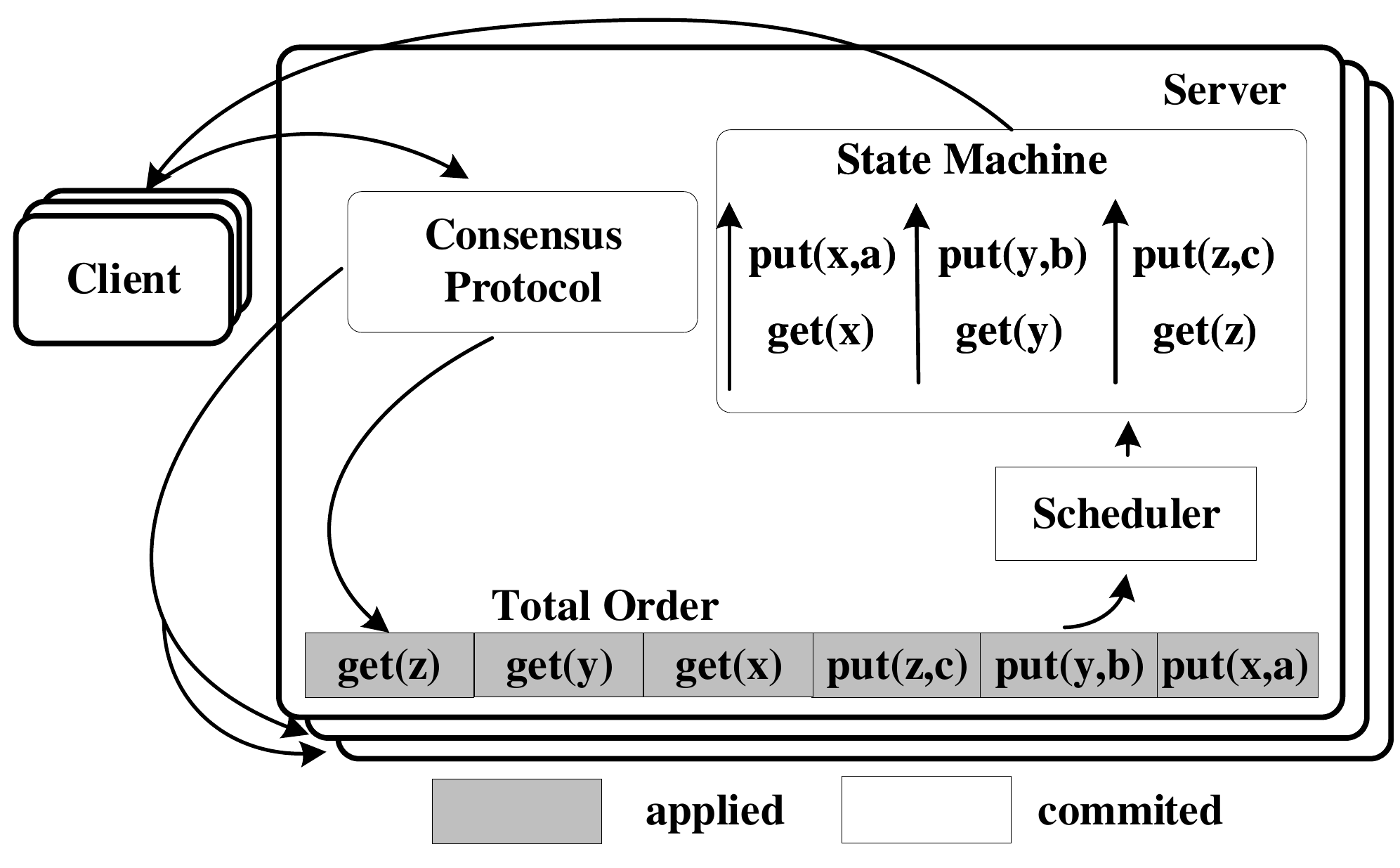}}
			\end{minipage}
		}
		\caption{Standard versus parallel state machine replication}
		\label{fig1}
	\end{figure}

In our system, the request messages are about transaction requests.
According to the Paxos protocol, each transaction has two states in a replica \emph{committed} and \emph{applied} (see Fig.~\ref{fig1}).
The \emph{committed} state represents that the transaction has been consistent with most of replicas but is not executed, and the state \emph{applied} represents that it has been executed in this replica.

\section{Parallel SMR} \label{Sec:ParallelSMR}
For the standard SMR, each replica executes the transactions sequentially following their total order (see Figure~\ref{fig1}\subref{fig-standard}).
Therefore, no matter how many thread resources are available, transactions can only be executed as if they were in a single-threaded replica.
With the development of high-speed networks and efficient consensus protocols (eg., \cite{lamport2006fast}\cite{marandi2010ring}), the CPU processing efficiency has becomes the next major performance bottleneck of SMR.
It is manifested by the fact that the speed of $applied$ is much slower than that of $committed$.
Although the concurrent execution of transactions causes uncertainty, the consistency will not be broken if only independent transactions are executed concurrently (see Figure~\ref{fig1}\subref{fig-parallel}).
As we know, two transactions are independent if they operate on different records or if they only read the same records.
For example, if a record is modified by one transaction, and operated by the other transaction, then these two transactions are said to be dependent or conflicted.
	
There have been some attempts (e.g. \cite{kotla2004high,alchieri2018early,7967165}) so far to boost SMR with parallel execution by exploiting such transaction dependencies.
In this section, CBASE\cite{kotla2004high} and its improved version batchCBASE\cite{7967165} are discussed, and conclusion of the motivation for our methods is presented in the end. More details about other related work can be found in Section \ref{Sec:RelatedWork}.
	
To parallelize the execution of transactions, CBASE sets up a scheduler for each replica.
The main part of the CBASE algorithm is shown in Algorithm \ref{Alg:CBASEScheduler}.
The core of the scheduler is a dependency graph, which takes transactions as vertexes and the dependencies among transactions as directed edges.
It keeps the partial order relationship (line 3) between transactions.
While accepting a transaction, the scheduler inserts it into the dependency graph (lines 6-8).
Based on the dependency graph, the scheduler dispatches free transactions to those idle threads (lines 18-19) in the thread pool for execution.
Once a transaction $t_i$ has been executed by a thread, the corresponding vertex and edges should be removed from the graph (line 20).
Thus other transactions without predecessor dependencies can be executed next.

	\begin{figure*}[htbp!]
		\subfloat[]{        \label{1fig2}
			\begin{minipage}[c]{.33\linewidth}
				\centering
				\centerline{\includegraphics[scale=0.6]{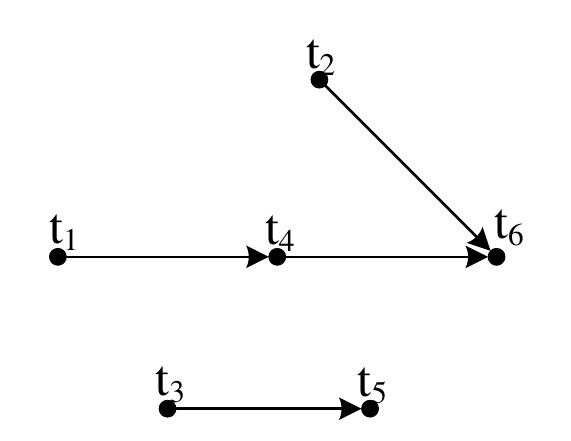}}
			\end{minipage}
		}
		\subfloat[]{      \label{2fig2}
			\begin{minipage}[c]{.33\linewidth}
				\centering
				\centerline{\includegraphics[scale=0.6]{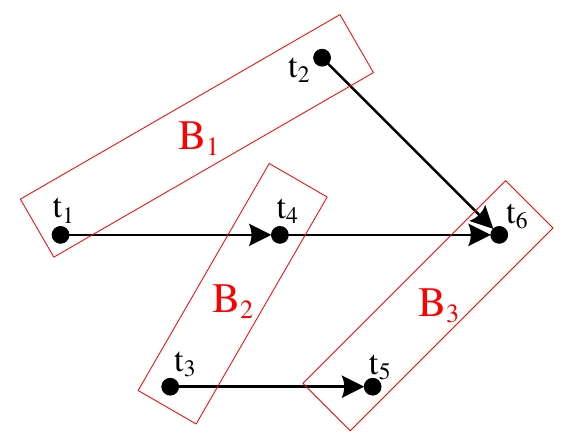}}
			\end{minipage}
		}
		\subfloat[]{        \label{3fig2}
			\begin{minipage}[c]{.33\linewidth}
				\centering
				\centerline{\includegraphics[scale=0.6]{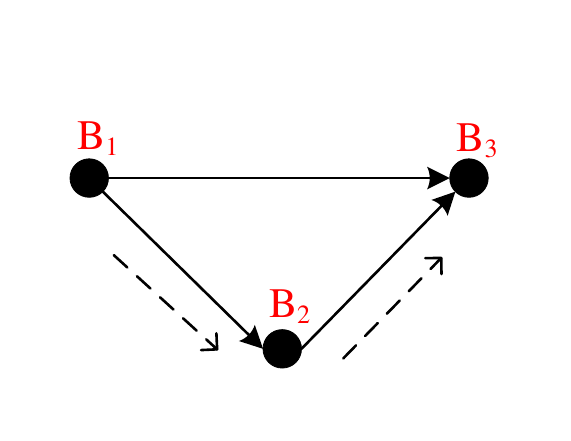}}
			\end{minipage}
		}
		\caption{ CBASE and batchCBASE dependency graphs. (a) the dependency graph of CBASE. (b) and (c) show how batchCBASE works, i.e., $t_1t_2$ becomes batch $B_1$, so they has to be executed sequentially though they are parallelizable; $t_2$ and $t_3$ have to be executed sequentially just because $t_1$ are conflict with $t_4$; Finally, all of them have to be executed sequentially as (c) where solid lines represents batch dependencies, and dotted lines represents execution trace.}
		\label{fig2}
	\end{figure*}

Figure~\ref{fig2}\subref{1fig2} shows how CBASE maintains the partial order of transactions based on the dependency graph.
These transactions are agreed at each replica in a total order sequence $[t_1t_2t_3t_4t_5t_6]$.
Among them, $t_1t_4t_6$, $t_3t_5$ and $t_2t_6$ are dependent subsequences.
For transactions in each such dependent subsequence, their position on the dependent graph path is determined by their relative order in the total order.
For example, $t_1 \rightarrow t_4 \rightarrow t_6$ represents that $t_6$ depends on $t_4$, and $t_4$ depends on $t_1$, because $t_1$ arrives $committed$ first, then $t_4$, and finally $t_6$.
New transactions need to be compared to all transactions in the graph to determine the dependencies.

Intuitively, the overhead of building a dependent graph is related to the number of nodes in the graph.
Specifically, the time complexity is $O(n^2)$.
Experiments in batchCBASE \cite{7967165} confirm that detecting conflicts (dependencies) between transactions is time consuming in heavy workloads.
Therefore, batchCBASE is designed to reduce the number of comparisons by packing transactions into batches, as the example shown in Figure~\ref{fig2}\subref{2fig2}.
Compared with CBASE, the detecting overhead of batchCBASE is reduced by a factor of the size of the batch.
Bitmaps technology is introduced to detect conflicts between batches.
It allocates a bitmap of 1,000Kbit for each batch.
If the intersection of two bitmaps is not empty, then it can be determined that the two corresponding batches have dependencies.
Therefore, the time complexity of batchCBASE dependency detection is $O(l(n/m)^2)$, where $l$ is a constant representing the time complexity of bit comparison using bitmap, $n$ is the number of transactions, and $m$ is the size of the batch.
However, such batch-based method has a higher the conflict probability between two batches.
In theory, the conflict probability between two random transactions is $1/n$, while the conflict probability between two batches is $p=\sum_{i=1}^m\binom{n}{i}(\frac{i}{n})^m(\frac{n-i}{n})^m$.
Thus, when the batch-based method is applied, the conflict probability has an exponential increase with respect to the batch size $m$.

Since transactions within each batch of batchCBASE is executed sequentially, the parallelism between transactions is reduced.
In addition, if any two conflicting transactions from each batches conflict with each, the two batches have to be executed sequentially as well because the two batches of transactions are considered to be conflicted in this case.
As shown in Figure~\ref{fig2}\subref{2fig2}, when the batch size is 2, it will degenerate into a sequential execution as Figure~\ref{fig2}\subref{3fig2}.

Moreover, since the scheduler internal operations of CBASE and batchCBASE, i.e. $dgInsertTrans(t_i)$, $dgRemoveTrans(t_i)$, and $dgGetTrans(t_i)$, are mutually exclusive, the call to any of these operations will lock the whole dependency graph (line 10,12) until it is finished.
From this perspective, the scheduler runs in a single-threaded mode, which introduces extra scheduling overhead.
	
	\renewcommand{\algorithmiccomment}[1]{\hfill\{#1\}}
	\makeatletter
	\def\BState{\State\hskip-\ALG@thistlm}
	\makeatother
	\begin{algorithm}[htbp]
		\caption{CBASE (bacthCBASE) scheduler}\label{Alg:CBASEScheduler}
		\begin{algorithmic}[1]
			\State \textbf{data structures and variables}
			\State \quad $Transaction$ \quad $t$ \Comment{or Batch $b$ for batchCBASE}
			\State \quad $DG = (T,E)$ \Comment{dependency graph, or $DG = (B,E)$}
			\State \quad ...
			\vspace{0.45em}
			\State \emph{The scheduler executes as follows}:
			\While {accept($t_i\in T$)} \Comment{or Batch $b_i$ for batchCBASE}
			\State \Call{dgInsertTrans}{$t_i$}
			\EndWhile
            \vspace{0.45em}
			\Procedure{dgInsertTrans}{$t_i$}
			\State $Lock(DG)$\Comment{occupy the whole DG}
            \State ...
			\State $Unlock(DG)$\Comment{release lock of DG}
			\EndProcedure
            \vspace{0.45em}
			\Procedure{T: dgGetTrans()}{}
            \State ... \Comment{omitted for simplicity}
			\EndProcedure
			\vspace{0.45em}
			\Procedure{dgRemoveTrans}{$t_i$}
            \State ... \Comment{similar to $dgInsertTrans()$}
			\EndProcedure
			\vspace{0.45em}
			\State \emph{Each worker thread executes as follows}:
			\While{$t_i \gets dgGetTran()$}
			\State execute transaction $t_i$\Comment{batchCBASE executes $b_i$}
			\State \Call{dgRemoveTrans}{$t_i$}
			\EndWhile
		\end{algorithmic}
	\end{algorithm}
	
To sum up, (i) CBASE has a greater overhead of detecting dependency; in addition (ii) batchCBASE increases the conflict probability which makes it highly likely to degenerate into sequential execution, (iii) as well as the running mode of their scheduler operations is single-threaded.
In our opinion, the main reason behind this is that the granularity of the scheduling object is not appropriate.
As stated earlier, transactions that access different records must be independent and need not be detected for the dependency.
As for CBASE, the scheduling granularity is the transaction which is so fine-grained that each dependency detection must be performed over all the other transactions.
As for batchCBASE, though the batch granularity is coarse enough to reduce comparisons, it still does not fully consider the dependency between transactions when selecting transactions to form a batch.
Therefore, it would be a better solution to organize transactions into a specific index structure according to the records to be accessed beforehand.
In this way, both efficient dependency detection and good parallelism can be achieved.

\section{Index-based Scheduler Model} \label{Sec:OurApproach}
We propose a deterministic and efficient parallel SMR scheduler for handling dependencies among transactions and scheduling them to execute concurrently on all worker threads available.
The proposed method dedicate to improving the performance of scheduler by designing a specific index structure and devising an elaborated concurrent scheduling scheme accordingly.
	
\subsection{Overall idea}
	The basic idea of the scheduler is as follows:
	\begin{itemize}
		\item The main part of the index structure is a simplified Bloom filter constructed from a single HashMap.
		Each key of the HashMap represents one record accessed by the transactions.
		Hence, without actually constructing and traversing the dependency graph, it can determine the dependency between transactions when they fall into the same Bloom filter bit by one hash. 
		\item The value corresponding to each key of the HashMap is a FIFO queue containing all the transactions accessing the record of the key.
		Hence, any different transactions at the heads of all transaction queues of the HashMap can be executed concurrently.
		\item Based on the above index structure, it is easy to make the scheduler concurrently perform scheduling operations (i.e., insert, remove, get) with record-granularity lock, which can guarantee safety and correctness as well.
	\end{itemize}

	\emph{\textbf{Transactions and Records:}}
	Transaction $t_i$ is composed of one or couples of commands and records.
We denote the total order of transaction $O_T$ as $(T, <_T)$ where $T=\{t_i|i=1,2...\}$ and $<_T$ represents the total order between two transactions.
	Let the transaction $t_i$'s record set $R_{t_i}=\{r_j| r_j \text{ is one of the record accessed by } t_i\text{'s commands}\}$ and the transaction set accessing the common record $r_j$ as $T_{r_j}=\{t_i| r_j \text{ is one of the records accessed by } t_i\text{'s commands}\}$.
	
	\emph{\textbf{Bloom Filter:}}
	The Bloom filter is constructed from a single HashMap.
	Although a Bloom filter is usually composed of more than one hash functions, the only one hash used here is the one of the HashMap.
	The reason is that our Bloom filter is used not only for testing the existence of dependencies but also for indexing transaction queues according to the record accessed.
	This is achieved by letting record $r$ be the key to be hashed and all the transactions in $T_r$ be the corresponding value mapped.
	Thus, for a transaction $t_i$, the time complexity of finding all dependent transactions related to record $r$ is O(1).
	
	\emph{\textbf{Transaction Queue:}}
	In order to provide efficient dependency detection and concurrent execution, all transactions in $T_r$ is organized in a FIFO (First In First Out) queue as the value part of our Bloom filter corresponding to the key $r$.
	The transaction queue $TQ_r$ of record $r$ is actually a relative order $O_{T_r}=(T_r,>_T) \subseteq O_T$.
	Thus, for a record $r$, the time complexity of inserting a transaction at the end of or removing a transaction from the head of the queue is O(1).
	Note that a transaction may exist in different transaction queue because it usually operate on multiple records.
	
\begin{figure}[htbp]
		\subfloat[]{        \label{afig3}
			\begin{minipage}[c]{.5\linewidth}
				\includegraphics[width=4.3cm,scale=0.4]{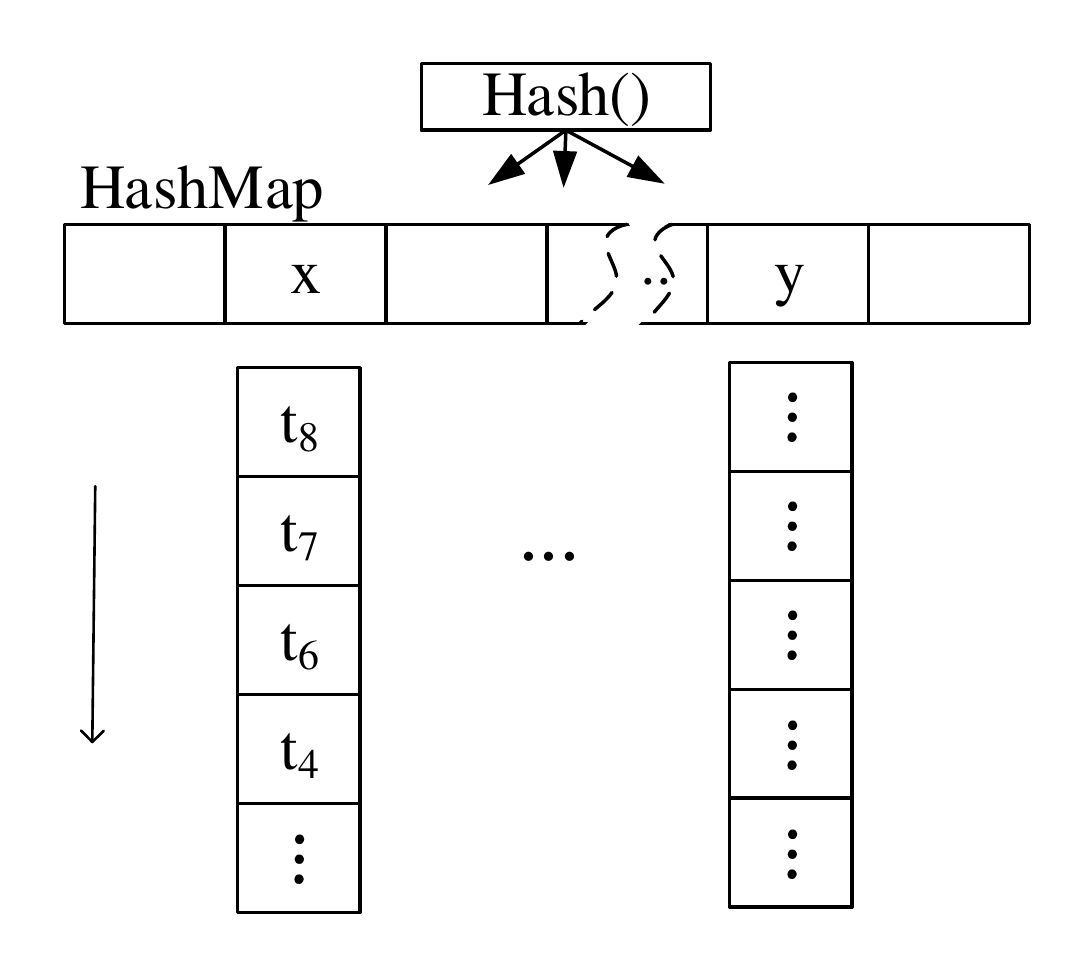}
			\end{minipage}
		}
		\subfloat[]{      \label{bfig3}
			\begin{minipage}[c]{.5\linewidth}
				\includegraphics[width=4.3cm,scale=0.4]{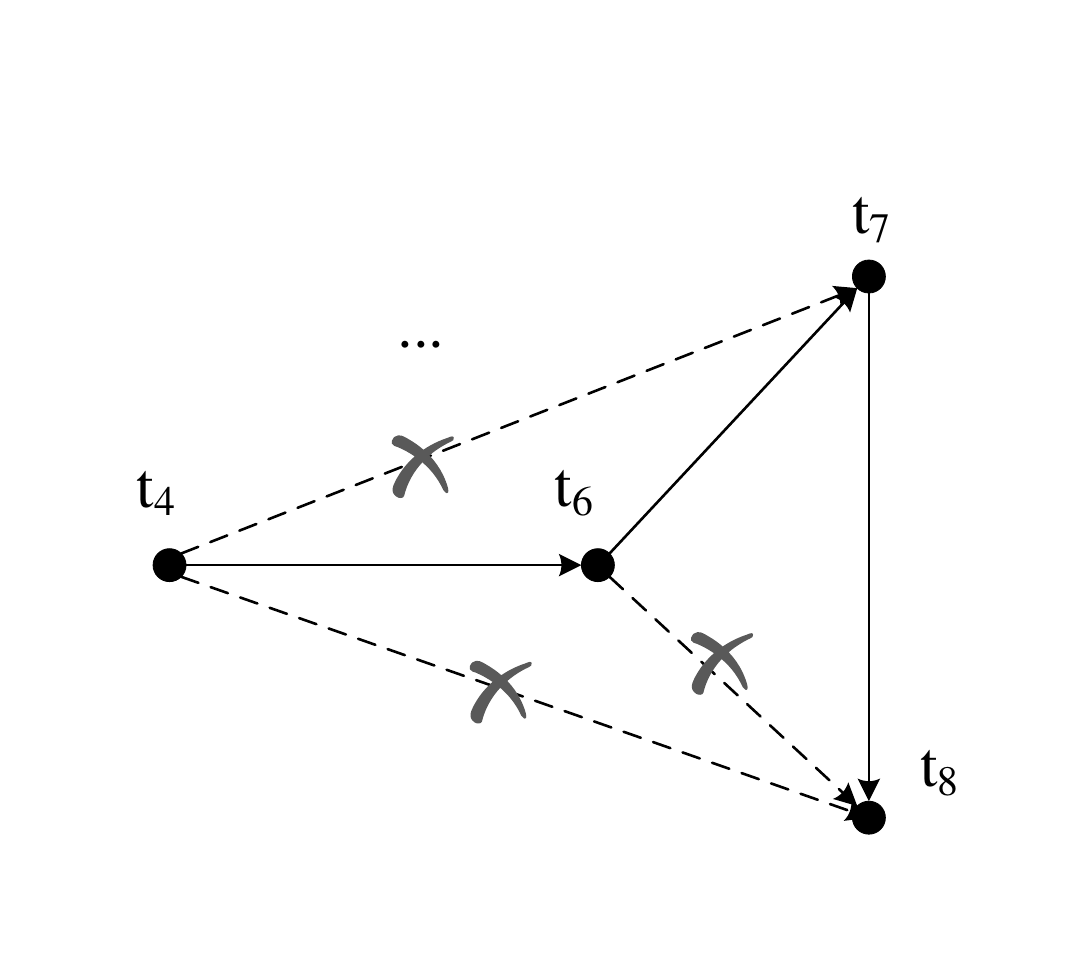}
			\end{minipage}
		}
		\caption{Example of index structure and simplified dependency graph. (a) The index structure with four transactions having the same records x in the same FIFO transaction queue. (b) The transaction queue in (a) only keeps necessary total order between every two adjacent transactions, eg. the edge $t_4$ to $t_7$, $t_4$ to $t_8$ and others in the origin dependency graph are omitted naturally.}
		\label{fig3}
	\end{figure}

	\emph{\textbf{Simplified Dependency Graph:}}
	All transaction queues together can form a simplified dependency graph which is consistent in order $<_T$ but much simpler in structure compared with the original complete dependency graph.Since the dependency relation and relative order between transactions are all transitive, it is not necessary to explicitly establish a complete total order through pairwise comparison within the transactions.
	Therefore, the proposed index structure can effectively reduce the overhead of detection and scheduling.
	Figure~\ref{fig3} exemplifies the basic idea of dependency graph simplifying.
	
	\emph{\textbf{Free Transaction:}}
    In our scheduler, after a transaction $t_i$ is inserted into scheduler, it is said to be \emph{\textbf{free}}, iff $\forall r_j \in R_{t_i}, TQ_{r_j}.head=t_i$.
    If a transaction is free, it can be scheduled to be executed.
If a transaction is still in transaction queue, it means that the transaction is under execution or not yet executed, in one word, unfinished.
	
	\emph{\textbf{Fine-grained lock:}}
	During the scheduling process, the granularity of operation locking is a record in our scheduler.
	It means that when the scheduler operates transaction $t_i$ on the above index structure, the scheduler will only lock those transaction queues corresponding to the records of $R_{t_i}$.
	Obviously, operations at the same location of the HashMap (i.e. transaction queues) are mutually exclusive, while operations at different locations are concurrent.
	Thus the maximum concurrency among these opeartions can be guaranteed.
	
	\subsection{Detailed algorithm}
	Algorithm \ref{Alg:OurScheduler} shows how our scheduler works in detail.
	The dependency graph is not explicitly defined because transaction queues can effectively replace it.
	When the system starts, procedure $Initialization()$ initializes a HashMap (line 7), and then initializes $N$ worker threads for waiting to execute transactions (lines 8-10).
	The length of HashMap does can be less than the number of records.
	In this case, there will be a certain probability that the hash function maps two different records to the same position.
	Fortunately, such false positives do not violate the consistency because those transactions that incorrectly fall into the same transaction queue will be safely executed sequentially.
	Although such a false positive transaction conflict may occur, it can guarantee that false negatives will never occur. 

	Once the scheduler accepts tranasctions, it will insert them into the index according to their total order (lines 12-14).
	As stated earlier, a transaction can be scheduled to be executed, if it does not depend on any other transactions, i.e., being \emph{free}.
	There are two situations.
	(i) For a newly accepted transaction, if there is no dependency detected, it can be executed directly after being inserted into the transaction queue after dependency detection;
	(ii) For a transaction in the transaction queue that has not been executed yet, it must be dependent and cannot be executed until its dependent transactions are all executed and removed.
	Therefore unlike CBASE and batchCBASE, our scheduler does not require a separate $dgGetTrans$ operation, but combines it with the insert operation and the remove operation to be $dgInsertAndGet$ and $dgRemoveAndGet$ respectively.
	They are detailed as follows:
	\makeatletter
	\def\BState{\State\hskip-\ALG@thistlm}
	\makeatother
	\begin{algorithm}[htb]
		\caption{Index-based  scheduler}\label{Alg:OurScheduler}
		\begin{algorithmic}[1]
			\renewcommand{\algorithmiccomment}[1]{\hfill\{#1\}}
			\State \textbf{data structures and variables}
			\State \quad $Transaction\ t$ \Comment{transaction}
			\State \quad $int\ N$\Comment{number of worker threads}
			\State \quad $TQueue\ TQ$\Comment{transaction queue}
			\State \quad $HashMap\ HM$\Comment{HashMap}
			\vspace{0.5em}
			\Procedure{Initialization()}{}
			\State initialize $HM$
			\State $\textit{N} \gets \text{desired number of worker threads }$
			\For {$id=1\dots N$}\Comment{initialize every worker thread}
			    \State create and start a worker thread $thr_{id}$
			\EndFor
			\EndProcedure
			\vspace{0.45em}
			\State \emph{The scheduler executes as follows}:
			\While {accept($t_i\in T$)}\Comment{accept $t_i$ from \textit{T}}
            \State $t_i.run = true$\Comment{used for $t_i$ executed exactly once}
			\State dgInsertAndGet($t_i$)\Comment{scheduler inserts $t_i$}
			\EndWhile
			\vspace{0.45em}
			\Function{bool: free}{$t_i$}
			\For{$r \in R_{t_i}$}
			\State  $TQ_r = HM(r)$\Comment{Bloom Filter used as index}
			\If{$t_i !=TQ_r.head$}
			\State  return $false$
		  	\EndIf
			\EndFor
			 return $true$
			\EndFunction
			\vspace{0.45em}
			\Procedure{dgInsertAndGet}{$t_i$}
			\For{$r \in R_{t_i}$}
            \State $TQ_r = HM(r)$
			\State  \Call{Lock}{$TQ_r$}
			\State $TQ_r.$\Call{insert}{$t_i$}
            \If{$r==R_{t_i}.last \land !free(t_i)$}
            	\State $t_i.run = false$
            \EndIf
            \State \Call{Unlock}{$TQ_r$}
			\EndFor
			\If{$t_i.run$}\Comment{no dependency afert insert}
			\State notify worker threads to execute $t_i$
			\EndIf
			\EndProcedure
			\vspace{0.45em}
			\Procedure{dgRemoveAndGet}{$t_i$}
			\For{$r \in R_{t_i}$}
			\State $TQ_r = HM(r)$\Comment{Bloom Filter used as index}
			\State \Call{Lock}{$TQ_r$}
			\State $TQ_r.$ \Call{remove}{$t_i$}
			\State \Call{Unlock}{$TQ_r$}
			\State $t_j = TQ_r.head$\Comment{candidate next to be executed}
			\If{$!t_j.run \land free(t_j)$}
                \State notify working threads to execute $t_j$
            \EndIf
			\EndFor
			\vspace{0.45em}
			\EndProcedure
			\State \emph{Each worker thread executes as follows}:
			\While{$t_i \gets \textit{notification from the scheduler}$}
			\State execute transaction $t_i$
			\State \Call{dgRemoveAndGet}{$t_i$}
			\EndWhile
		\end{algorithmic}
	\end{algorithm}
	
	\emph{\textbf{dgInsertAndGet:}}
	The transaction's execution order of $dgInsertAndGet(t_i)$ and $dgInsertAndGet(t_j)$ are subject to the order of $t_i$ and $t_j$ in $O_T$.
	Therefore, they can not run concurrently.
	A call to $dgInsertAndGet(t_i)$ consists of two operations, i.e., the operation of inserting $t_i$ into those transaction queues that correspond to each record $r \in R_{t_i}$ (lines 22-24), and the operation of determining whether $t_i$ can be executed (lines 25-29) now.
	According to previous description of \emph{Free Transaction}, if $t_i$ appears at the head of all corresponding transaction queues after insertion, it must be free and can be executed immediately because $t_i$ is the only transaction in those queues.
	More intuitively speaking, it has no dependent incoming edges in the dependency graph.
	Otherwise, $t_i$ can not be executed directly.
	Thus it will be scheduled to worker threads in $dgRemoveAndGet(t_j)$.
	The lock granularity of the operation is a record (line 23 and 27), i.e. only one transaction queue corresponding to each record in $R_{t_i}$ is locked at a time.
	It does not lock all transaction records at the same time, ensuring maximum concurrency with operation $dgRemoveAndGet(t_j)$.
	
	\emph{\textbf{dgRemoveAndGet:}}
	Just like $dgInsertAndGet$, removing a finished transaction $t_i$ from the index also needs to operate on multiple transaction queues.
	With the help of HashMap in our index, those transaction queues that correspond to each record $r \in R_{t_i}$ can be easily obtained (line 32).
	In our scheduler, transactions to be executed or finished transactions to be removed are kept at the head of corresponding transaction queues, which makes the remove operation more efficient.
	According to previous description of \emph{Free Transaction}, if a transaction $t_i$ is free, it must appear at the head of those transaction queues $TQ_r$ where record $r\in R_{t_i}$.
	Thus transactions at the head of each $TQ_r$ is checked for free after removing finished transaction $t_i$.
	Then free transactions can be dispatched to available worker threads for executing next.
	The execution checking is safe and does not need to acquire locks of other transaction queues.
	Both $dgInsertAndGet$ and $dgRemoveAndGet$ achieve the goal of not having to lock all transaction queues.
	The operations of index-based scheduler have the maximum concurrency when it is measured by the number and granularity of the lock.

	\subsection{Correctness}
	The key to the design of the scheduler is to ensure the security of scheduling operations and the consistency of the state of the transaction execution results between replicas.
	Here, we will highlight the security of deadlock-free and hungry-free and the validity of replica consistency of our index-based scheduler from data structure, lock granularity and scheduling strategy.
	See the appendix for details.
	\begin{itemize}
		\item [1)] \textbf{Operation safety:}
		In the case of fine-grained locks, scheduler is deadlock-free and hungry-free.
		i) Deadlock-free.
		First of all, for any two transactions $t_i$ and $t_j$, the index-based scheduler will never produce scheduling results where their operations dependent on each other.
		Both $dgInsertAndGet()$ and $dgRemoveAndGet()$ are required to be executed sequentially in FIFO order.
		Thus deadlocks never occur;
		ii) Hungry-free.
		During $dgInsertAndGet$ operation, all free transactions are scheduled to be executed directly by the scheduler.
		Non-free transactions met in this operation can only turn free during the following $dgRemoveAndGet$ operations in which those transactions they depend on are executed and removed.
		Hence, the transaction scheduling process is always driven by insertion and deletion operation.
		As long as there are unexecuted transactions in the transaction queue, they will eventually be executed.
		Thus hunger never occurs.
		
		\item [2)] \textbf{Replica consistency:}
	    The transaction queue can ensure $<_T$ order between transactions.
		Although only $<_T$ order between any two adjacent transactions are maintained, the transaction queues are still consistent with the complete dependency graph w.r.t $<_T$ order.
		Suppose there is competition for locks, and $dgInsertAndGet(t_i)$ and $dgRemoveAndGet(t_j)$ operate on the same records $r$ 
		in the same transaction queue $TQ_r$.
		No matter which operation executes first, it will not affect the $<_T$ order between each transactions.
		In addition, it is important for a transaction to execute only once in order to keep replica consistency. 
		In order to determine whether a transaction can be executed in both insert and deletion operations, a flag $run$ is defined in the transaction $t_i$, it can guarantee $t_i$ executed exactly once even when $t_i$(line 39) appears in both $dgRemoveAndGet$ and $dgInsertAndGet$. From the system perspective, Paxos protocol guarantees the unique total order between replicas on which the same index-based schedulers run as described in Algorithm \ref{Alg:OurScheduler}, even though the execution speed of each replica may be different, the related records of replicas will reach the same states whenever a transaction $t_i$ become $applied$.
	\end{itemize}
	
\section{Experiments}\label{Sec:Experiments}
This section will introduce the system prototype, experiment configuration, experiment purpose, experiment method and conclusion of our experiment.
\subsection{System prototype}
To evaluate the performance of our index-based scheduler, called fastCBASE, we implemented an in-memory database in C/S service model.
This system provides three transaction operations: PUT, GET and DELETE.
Algorithm CBASE, batchCBASE, and our fastCBASE are all implemented on it with different schedulers.
Clients sequentially send transaction commands, and the replicas first agree on a total order of all transactions received, and then the corresponding operations are performed by the specific scheduler.
The implementation of Algorithm \ref{Alg:CBASEScheduler} follows \cite{7967165}.
We published the source code of Algorithm \ref{Alg:CBASEScheduler} and our Algorithm \ref{Alg:OurScheduler} online\cite{fastcbasewebsite}.

\subsection{Environment}
Our experimental environment consists of a cluster of four HP nodes. Three of them work as servers, playing the role of proposer and acceptor in Paxos protocol, and each has 2 E5-2620 CPU, 2.10GHz, hyper-threading, a total of 24 threads, and 256G memory.
The client is deployed in the other HP node which has a four-way E7-4820 CPU, 2.0GHz, 8 cores per channel, hyper-threading, a total of 64 threads.
The operating systems are all Ubuntu 18.04.2 LTS.
The clients in the client node send large number of transactions to make the servers fully loaded.
All applications are implemented by Go language version go1.12.1.
The communication within the cluster goes through ER3200G2, a gigabit network switch.

\subsection{Goals and methods}
Since the index-based scheduler is proposed to ensure the maximum concurrency among transactions as well as a lower scheduler load, the main experimental purposes include:
	
	\begin{itemize}
		\item the speed-up achieved compared to state-of-art
		\item the scalability with a growing number of worker threads
        \item the impacts of scheduling overhead
		\item the false positive introduced by Bloom filter
		\item the impacts of conflicts on performance of scheduler
	\end{itemize}

For the first point, in order to compare the performance of our scheduler with other schedulers, we evaluate each scheduler's performance under conflict-free workloads, and compare the performance under the same number of worker threads with CBASE and batchCBASE.
	
For the second point, we evaluate the performance improvement of our scheduler with an increasing number of threads under the conflict-free workloads, and compare it with CBASE and batchBASE.
	
For the third point, we can analyze with the above experimental results.
	
For the fourth point, since batchCBASE uses two bitmaps bitwise comparison methods in conflict detection and our scheduler uses Bloom Filter, all of them will introduce false positive conflict. We compare the false positive rate introduced by these two scheduler models under different bitmap(HashMap) sizes.
	
For the fifth points, we compare the performance changes of our scheduler and batchCBASE under different conflict rate workloads.
	
	\subsection{Speed-up analysis}
	\begin{figure}[htbp]
		\centerline{\includegraphics[scale=1]{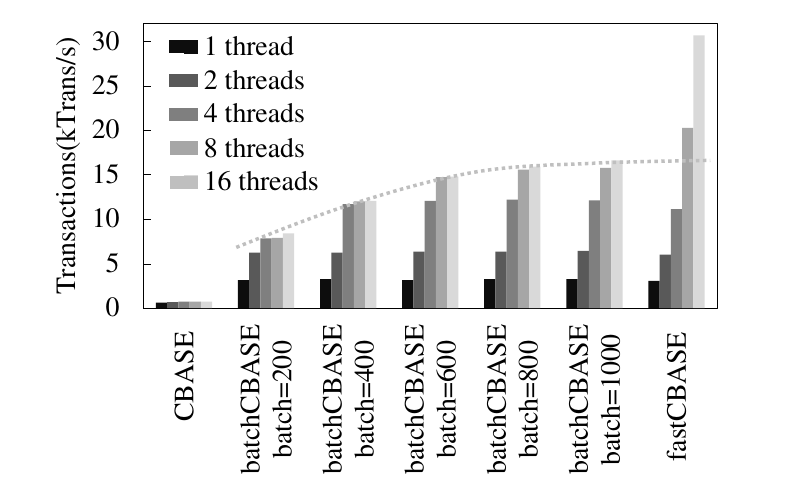}}
		\caption{Threads scalability for contention-free workloads}
		\label{excel1}
	\end{figure}
In order to observe the most obvious speed-up ability of each scheduler, we analyze the throughput of each scheduler in the case of workloads without conflicting transactions, which is indicated by an average throughput each replica per second. Figure~\ref{excel1} shows the system throughput of CBASE, batchCBASE and our fastCBASE without conflict. The performance of different batch sizes are tested since the batch size of batchCBASE has a significant impact on it.

It can be seen that the traditional CBASE has a very low performance, because the scheduler has a large overhead in the dependency detection, which severely limits the throughput of the whole system.
The speed-up of CBASE is poor.
As the number of worker threads increases, its performance does not increase significantly.
With 16 threads, it only achieves a throughput of about 1000Trans/S. Even though many worker threads are avaliable, the scheduler cannot fully utilize them.

To improve the performance of the system, batchCBASE sacrifices scheduling freedom (transaction sequential execution within batch) for a lower number of comparisons.
It can be seen that the performance of batchCBASE is more significantly improved than that of CBASE, which effectively solves the problem of heavy scheduling workload in CBASE.

As can be seen from Figure~\ref{excel1}, the performance of batchCBASE increases linearly with the increase of the number of threads in 1, 2, 4, but it does not increase with batch size, which indicates that neither one of the schedulers is a performance bottleneck for the system, throughput is limited by the number of available worker threads.

Our fastCBASE and  batchCBASE have a similar performance at 1, 2 threads, fastCBASE has a slightly lower performance than batchCBASE when the number of threads is 4.
That is because compared to the batch method, in order to ensure the concurrency between each two transactions, the operating system needs to allocate transactions to worker threads in the granularity of single transaction, whereas batchCBASE executes in the granularity of single batch, there is no need for frequent scheduling within a single batch.
Under the conflict free workload, when the number of available threads of the system increases, the operating system requires a higher thread synchronization overhead compared to a less number of threads.
However, as the number of threads increases, the performance gain of our method is much higher than that of batchCBASE.
	
While reducing the overhead of comparison, batchCBASE also potentially reduces the scheduling load.
For example, for 10,000 transactions, when the batch size is 1000, there are only 10 batches in the scheduler, and the scheduling load is relatively low (so for batchCBASE, the scheduling load of procedures is not the performance bottleneck of the scheduler).
The larger the batch of batchCBASE is, the lower the scheduler workload and the better the performance will be.
The speedup of its performance gradually stabilizes gradually as the batch increases, as shown in Figure~\ref{excel1}.
By the start of 8 threads, the performance no longer increases linearly with the number of threads in every batch.
This is because although the number of comparisons is reduced by batch, the load scheduling is still relatively high compared to our scheduler.

Because of our elaborate concurrent scheduling process based on the special index structure, although our scheduler needs to manage each transaction, it is still even more efficient than batchCBASE.
Figure~\ref{excel1} shows that the throughput of our scheduler in 8 and 16 threads is much higher than batchCBASE.
And unlike batchCBASE, the performance of our method improves much near linearly with the increase of the number of threads, so it has strong scalability.

	\subsection{Conflict rate analysis}
Dependency detection based on the index structure of fastCBASE becomes simple and efficient.
If the records corresponding to the new transaction conflict after hashing with HashMap, it means that there has been the transaction containing the record in the index.
However, if the number of records is greater than the length of the HashMap, then the Bloom Filter may have false positive, i.e. different records will be mapped to the same location in the HashMap.
However, our Bloom Filter is also used as index, so it can only be set up with one hash function.
Each batch in batchCBAse corresponds to a bitmap.
When comparing two batches, the bitmap is compared by bitwise comparison to determine whether there is conflict, so the false positive may also be generated.
We compare the conflict rate generated by our fastCBASE scheduler with Bloom filter and compare the rate generated by batchCBASE with the  bitmap of each batch.
	
Before evaluating the conflict rate, recall mathematical formula of conflict rate proposed in Section \ref{Sec:ParallelSMR}.
When the batch is not executed, the probability of conflict between batches in batchCBASE is an exponential times with respect to batch size $m$.
Although the conflict rate can be accurately represented, the result is not intuitive, hence simulation experiments for different schedulers are performed.
	
In the simulation, unfinished transactions in the scheduler are stored in the transaction queue.
For our scheduler, the new transaction generated by the simulation is detected by a Bloom Filter(the transaction queue and the Bloom Filter used in simulation differ from the previous).
The number at each location of the Bloom Filter represents the number of its conflict transactions.
If the corresponding Bloom Filter's location of the newly generated transaction is not zero, the conflict is considered. When the conflict detection is completed, the oldest transaction in the transaction queue is removed, the corresponding location of the Bloom filter is reduced by 1, and the new transaction is added to the transaction queue.
If the new transaction conflicts with all transactions, the conflict rate is 100\%. If it does not conflict with any transactions, the conflict rate is 0.
Therefore, the conflict rate can be defined as: the conflict proportion of the new transaction and the unfinished transactions in the queue at a given period of time or at a specific length of the queue.
In our simulation, we use a fixed length of transaction queue to calculate conflict rate.  For batchCBASE, only the conflict detecion is different. If at least one common bitmap position
is set as 1 in both bitmaps of the two batches, then a conflict is computed.
	
In our simulation experiment, a transaction contains only one record.
We randomly generate $10^8$ records.
Thus the probability of generating the same record twice in the simulation is almost zero ($10^{-8}$), which means conflict rate generated is mainly caused by false positive.
In our scheduler, the impact on the conflict rate mainly comes from the size of HashMap.
The conflict rate of batchCBASE is also affected by the size of bitmap and batch.
We conducted $10^6$ times simulation, the length of the transaction queue is set to 10,000, ie., there are average 10,000 unfinished transactions in the scheduler.
The corresponding batchCBASE has a graph size of 50 nodes when the batch size is 200, and a graph size of 25 nodes when the batch size is 400.
And we set up the HashMap and bitmap size to be 100K and 1M respectively. The experimental results are show in Table~\ref{tab1}.
	\newcommand{\tabincell}[2]{\begin{tabular}{@{}#1@{}}#2\end{tabular}}
	\setlength{\tabcolsep}{0.7mm}{
		\begin{table}[htbp]
			\caption{Conflict rate}
			\begin{center}
				\begin{tabular}{|c|c|c|c|}
					\hline
					\tabincell{c}{\textbf{HashMap}\\\textbf{size}}&\tabincell{c}{\textbf{fastCBASE conlict}\\\textbf{rate}}&\tabincell{c}{\textbf{batchCBASE conlict}\\\textbf{rate,batch=200}}&\tabincell{c}{\textbf{batchCBASE conlict}\\\textbf{rate,batch=400}}\\
					\hline
					102400&0.000984\%&32.558\%&79.332\%\\
					\hline
					1024000&0.0000975\%&3.844\%&14.796\%\\
					\hline
				\end{tabular}
				\label{tab1}
			\end{center}
		\end{table}
	}
	
It can be seen from Table~\ref{tab1} that under the same configuration, the conflict rate of batchCBASE is nearly 10,000 times of the rate of fastCBASE, which meets the expectations of the mentioned mathematical formula.
As the size of HashMap or bitmap increases, the conflict rate of fastCBASE and batchCBASE will decrease, but batchCBase will amplify the conflict rate due to batch, which will also increase the false positive rate.
Therefore, in reality, even if the conflict rate is very low, batchCBASE will still be greatly affected, while the false positive rate brought by our scheduler would hardly affect the performance.
Next experiments will confirm these two aspects.
	
	\subsection{Speed-up analysis for conflict-prone workloads}
	\begin{figure}[htbp]
		\centerline{\includegraphics[scale=1]{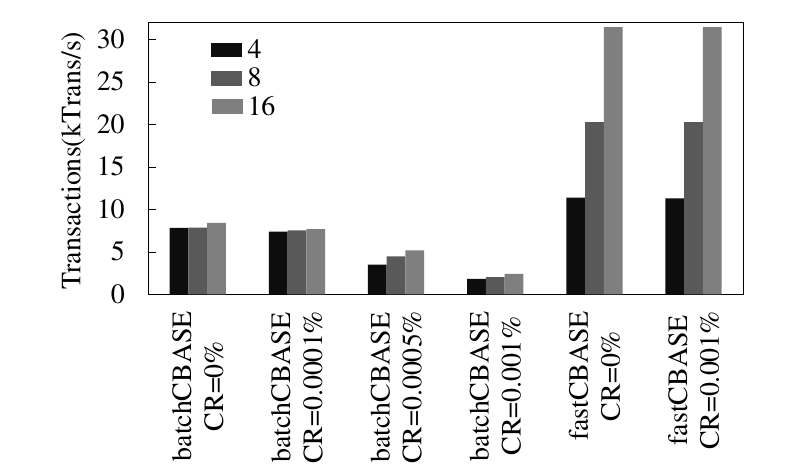}}
		\caption{Throughput under light conflict rate workload}
		\label{excel2}
	\end{figure}
	\begin{figure}[htbp]
		\centerline{\includegraphics[scale=1,height=4.2cm,width=7.8cm]{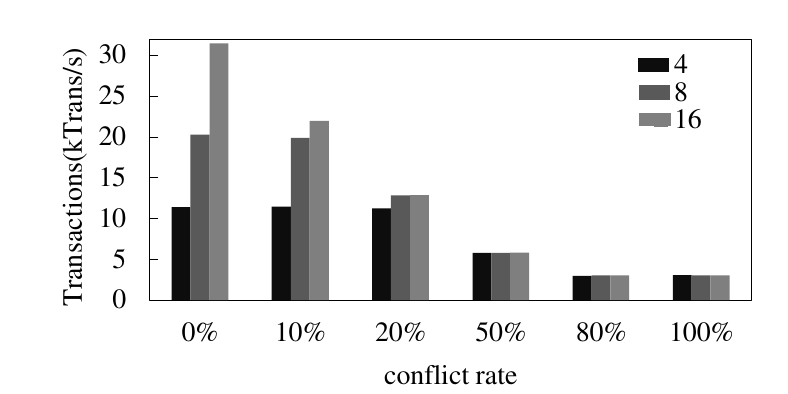}}
		\caption{Throughput under different conflict rate workloads}
		\label{excel3}
	\end{figure}
Now we analyze the performance of our scheduler and batchCBASE as the conflict rate increases.
According to the analysis in the previous section, the configuration that leads to the lowest false positive rate is adopted, that is, the lengths of HashMap of fastCBASE and bitmap of batchCBASE are 1M, and the batch size of batchCBASE is 200.
According to the analysis of the conflict rate in the previous section, although the actual conflict rate is extremely low, batch method will still amplify its effect.
In addition to the sequential execution of the internal transactions of the batch, the decrease of its performance is significantly magnified.
As Figure~\ref{excel2} verifies, a slight increase in the conflict rate of batchCBASE causes a drastic drop in the performance.
In contrast, the fastCBASE scheduler is not change significantly affected the corresponding conflict rate.

Figure~\ref{excel3} shows that the throughput of our scheduler decreases with the increase of conflict rate.
When the conflict rate is 10\%, only the throughput of 16 threads decreases.
This is because with the increase of the conflict rate, the parallelism of transaction execution decreases, consequently the utilization of multi-threading is reduced.
For the same reason, when the conflict rate is 20\%, the throughput of 8 threads begins to decrease.
And as the conflict rate continuous to increase, the performance gain caused by the increase of the threads' number is significantly reduced.
When the conflict rate is more than 50\%, i.e. more than half of the transactions cannot be executed in parallel, the redundant threads cannot be utilized, and the performance on different threads is approximately equal.

Based on the results of Figure~\ref{excel2} and Figure~\ref{excel3}, it can be known that our scheduler can allow a maximum parallelism among transactions.
When the conflict rate reaches 20\%, there is still similar performance to the batchCBASE under conflict-free workload.
With the conflict rate increasing, our scheduler is more robust.
	
	\section{Related Work}\label{Sec:RelatedWork}
CBASE\cite{kotla2004high} and batchCBASE\cite{7967165} propose to set up a deterministic scheduler on the replica, and actually they are late scheduling model\cite{8671580}, about which more details have been described above. Different from CBASE, an early schedule system model is proposed by \cite{marandi2014rethinking}. By setting up a client proxy, all client transactions are grouped according to transaction's semantic. Independent transactions can be allocated to different groups, and dependent transactions must be allocated to the same group. Each group of transactions is sent to all servers by atomic broadcast. Serve-side proxy maps groups to specific threads. A transaction may conflict with transactions in multiple groups. Therefore, synchronization among groups is required to ensure that the transaction is executed only once. To optimize the process of thread scheduling in\cite{marandi2014rethinking}, a multi-objective programming model\cite{alchieri2018early} is proposed to maximize parallelism and minimize execution time. The constraint is to ensure the relative order among the transactions. In order to achieve the optimal scheduling results, high time complexity is required either. Therefore, the existence (or absence) of an optimization model that combines early scheduling and concurrency is still an open question.
	
Eve\cite{evekapritsos2012all} implements deterministic parallelism through a scheduler called mixer, which groups requests into batches, and replicas execute batch transactions in parallel in a speculative manner. After the batch execution, the validity of the replica status is checked during the validation phase. If too many replicas are inconsistent, the replica will roll back to the previous validated state and re-execute the command sequentially. Eve is therefore a process of execute-validation. Unlike Eve, Storyboard\cite{kapitza2010storyboard} enhances SMR through a prediction mechanism that predicts the order of locks across replicas based on application-specific knowledge. When the prediction is correct, the transactions can be executed in parallel. If the prediction does not match the execution path of the transactions, the replica must establish a deterministic execution sequence with other replicas through consensus protocol. In this case, Storyboard stops the current execution and repredicts the execution path. All replicas will re-execute the transactions based on the new path.

Rex\cite{guo2014rex}, an execute-agree-follow model, in which a primary machine is free to execute transactions concurrently at first, and uncertain decisions are recorded in a partial order trace, and then other secondary machines will receive the trace. Finally, the secondary machine executes the same trace concurrently, which keeps consistency with the primary machine. Rex detects the relationship of the transactions by transaction-to-lock competition, encapsulates this detection mechanism into c++ synchronization primitives, so any applications developed with this synchronization primitive can generate trace. Traces can be generated only after the transactions on the primary server being executed, so replication on the secondary is a passive replication process, requiring a higher reconfiguration cost when the primary downtime occurs.

\section{Conclusion}\label{Sec:Conclusion}
In order to promise a high performance, the parallel state machine replication requires an elaborated design to execute independent transactions in parallel and dependent transactions following their relative order. To achieve this goal, efficient and correct dependency detection and scheduling strategies are needed. The existing models cannot make a good balance in these aspects, their advantages lead to their weakness, so their scheduler is inclined to become the performance bottleneck of the system. In this paper, an efficient scheduler based on a specific index structure is designed to detect dependency, express partial order relations and to schedule transactions, which can ensure the maximum parallelism of the execution between transactions to fully exploit the advantages of multi-core processors, and also can keep consistency among replicas.
	
	\bibliographystyle{./IEEEtran}
	\bibliography{./conference}

	\section*{APPENDIX}

\begin{definition}[Total order]
A transaction sequence is a pair($T$,$<_T$) where $T$ is a set of transactions and $<_T \subseteq T \times T$ is an irreflexive and antisymmetric total order (this total order represents Paxos functionality)
\end{definition}

\begin{definition}[Conflict, dependency relation]
Tow transactions $t_i$ and $t_j$ conflict if $R_{t_i} \cap R_{t_j} \neq \varnothing$.
Given a conflict relation $\# T \subseteq C \times C$ among transactions, the dependency relation $\vdash$ is the transitive closure of $<_T \cap \# T$, so it is an partial order.
\end{definition}
	


\begin{definition}[Dependency Graph]\label{Def:DG}
Given a new transaction $t_i$, the dependency graph $DG=(T,E)=TQ,TQ=\{TQ_r|\forall r \in R_{t_i},i=1,2...\}$, where $T=\{t_j|\forall t_j \in TQ,j=1,2...\}$ and every two adjacent transactions $t_i$ and $t_j$ in the same $TQ_p$ have the relation of $t_i \vdash t_j$, so by construction $\vdash$ is equivalent to the edges $E$ in $DG$.
\end{definition}


    \noindent\textbf{Replica consistency 1.}
    Transaction is executed exactly once.
    Suppose the $R_{t_i}$ of $t_i$ is $\{x_1,x_2...x_l\}$ and the $R_{t_j}$ of $T_j$ is $\{y_1,y_2...y_m\}$, there exits some records $\{x_{i1},x_{i2}...x_{in}\}$ equal to $\{y_{j1},y_{j2}...y_{jn}\}$ and in corresponding transaction queues $t_i=t_j.prior$ (when the remove operation of $t_j$ is over, now $t_j$.prior is the head of $TQ$(line 39)).
    At this time, $t_i$ and $t_j$ detected whether to be executed in these two operations are actually the same transaction.

    We define $dgInsertAndGet$=($\lfloor \textcircled{1},\textcircled{2}  \rceil$), where \textcircled{1} represents the insert operation (lines 22-24, Algorithm 2, the same as folloing), \textcircled{2} represents the detection operation (lines 25-29) after insert, and ``$\lfloor \rceil$'' means all operations within it are protected by lock(line 23,27 and line 33,35);
    Similarly, $dgRemoveAndGet$=($\lfloor \textcircled{3}\rceil,\textcircled{4}$), where \textcircled{3} represents the remove operation (lines 32-35), and \textcircled{4} represents the detection (lines 36-38) after remove.
    The detection operation \textcircled{2} of $dgInsertAndGet$ happens after performing operation \textcircled{1} on the last record $x_l$ of $R_{t_i}$.
    If detection operation \textcircled{4} in $dgRemoveAndGet$ success on the transaction queue corresponding to record $y_{jn}$, the former detection operation in $dgInsertAndGet$ must be failed because the remove operation for $t_j$ on later transaction queue like $y_{jn}$ has not been executed yet.
    In this most extreme case, all possible order of processing are ($\lfloor \textcircled{1},\textcircled{2} \rceil,\lfloor\textcircled{3} \rceil,\textcircled{4}$), ($\lfloor\textcircled{3}\rceil
,\textcircled{4},\lfloor \textcircled{1},\textcircled{2}\rceil$
), ($\lfloor \textcircled{3}\rceil
,\lfloor\textcircled{1},\textcircled{4},\textcircled{2}\rceil
$), ($\lfloor\textcircled{3}\rceil
,\lfloor \textcircled{1},\textcircled{2},\textcircled{4}\rceil$
).
    According to Algorithm \ref{Alg:OurScheduler}, it can promise only \textcircled{2} will schedule $t_i$ to be executed.
    In other cases, the safety can be guarenteed following the fact given below.

\vspace{0.3em}
	\noindent\textbf{Replica consistency 2.} The dependency graph (DG) is a directed acyclic graph (DAG).
	According to Definition \ref{Def:DG}, $TQ$ can keep the relative order between dependent transactions, i.e. $\vdash$.
	Since $\vdash$ is a partial order, $DG$ is a $DAG$.

\vspace{0.3em}
	\noindent\textbf{Operation safety 1.} No deadlock.
	According to Replica consistency 1, $TQ$ is acyclic because there is no transactions in $TQ$ dependent on each other.
	Since both $dgInsertAndGet$ and $dgRemoveAndGet$ operate the records in $TQ$ in FIFO order, there always exits transaction at the head of it's corresponding $TQ_r$ as long as $TO$ is not empty, which means that it does not depend on others and is free to be executed.

\vspace{0.3em}
	\noindent\textbf{Operation safety 2.} No starvation.
	$dgInsertAndGet(t_i)$ inserts the transactions at the end of queues, and detection of $dgRemoveAndGet(t_j)$ checks the transactions at the head.
	Based on Replica consistency 1, if $t_i \neq t_j.prior$, the detection of $t_j.prior$ will be executed by operation \textcircled{4}.
	As for $t_i$, in operation \textcircled{2}, the flag $run_i$ will be set false, so the detection of whether $t_i$ can be executed (line 37).
	As each transaction $t_i$ has an order in $<_T$ and no deadlock, $t_i$ will be executed eventually.

\vspace{0.3em}
\noindent\textbf{Replica consistency 3.}
Since $O_T$ is the same for all replicas, and the execution is subject to $\vdash$ of the dependent transactions, there is no deadlock and no starvation.
All transactions will be executed exactly once, thus all the replicas will have the same identical states after every transaction in $T$ is finished.
\end{document}